\documentclass [12pt,preprint]{aastex}
\usepackage{epsfig}
\begin{document}

\voffset-0.5cm
\newcommand{\gsim}{\hbox{\rlap{$^>$}$_\sim$}}
\newcommand{\lsim}{\hbox{\rlap{$^<$}$_\sim$}}

\title{On The  Recently Discovered Correlations Between\\  
Gamma-ray And X-ray Properties Of Gamma Ray Bursts}

\author{Shlomo Dado\altaffilmark{1} and Arnon Dar\altaffilmark{1}}

\altaffiltext{1}{Physics Department, Technion, Haifa 32000, Israel}

\begin{abstract}

Recently, many correlations between the prompt $\gamma$-ray emission 
properties and the X-ray afterglow properties of gamma-ray bursts (GRBs) 
have been inferred from a comprehensive analysis of the X-ray light curves 
of more than 650 GRBs measured with the Swift X-ray telescope (Swift/XRT) 
during the years 2004-2010. We show that these correlations are predicted 
by the cannonball (CB)  model of GRBs. They result from the dependence of 
GRB observables on the bulk motion Lorentz factor and viewing angle of the 
jet of highly relativistic plasmoids (CBs) that produces the observed 
radiations by interaction with the medium through which it propagates. 
Moreover, despite their different physical origins, long GRBs (LGRBs) and 
short hard bursts (SHBs) in the CB model share similar kinematic 
correlations, which can be combined into triple correlations satisfied 
by both LGRBs and SHBs.

\end{abstract} 


\section{Introduction}

As the measurements of the properties of gamma ray bursts 
(GRBs) and their afterglows (AGs) become more 
complete and more accurate, new correlations between these properties  are 
being discovered. Such correlations challenge GRB models. Recently, 
Margutti et al.~(2013) reported results from a comprehensive statistical 
analysis of the X-ray light-curves of more than 650 GRBs
(GRBs) measured with the Swift X-ray telescope (Swift/XRT) between 2004 
and the end of 2010. In particular, they reported the discovery of new 
correlations between the properties of the $\gamma$-ray and the X-ray 
emission in GRBs. In this paper, we show that the newly discovered 
correlations by Margutti et al.~(2013) between the $\gamma$-ray and X-ray 
properties of GRBs agree well with those predicted by the 
cannonball (CB) model of GRBs (Dar \& De R\'ujula~2004, Dado et 
al.~2009a,b and references therein). Like the previously well established 
correlations between GRB observables, in the CB model they all have the 
same simple 
kinematic origin - they result from the strong dependence of the observed 
radiations on the bulk motion Lorentz and Doppler (L\&D) factors of the 
jet of 
highly relativistic plasmoids (CBs) that produces the radiations by its 
interaction with the medium through which it propagates (Dar and De 
R\'ujula~2000,2004, Dado et al.~2007, Dado and Dar~2012a).

In this paper, we use the CB model dependence of the $\gamma$-ray and 
X-ray 
properties of GRBs on the Lorentz and Doppler factors to derive the 
kinematic correlations between them. For completeness, we shall first 
summarize the origin of the observed radiations in the CB model and the 
expected correlations among the properties of the prompt $\gamma$-ray 
emission and among the properties of the X-ray afterglow before 
proceeding to derive the correlations between them.

\section{Origin of the observed radiations in the CB model}
In the cannonball (CB) model of GRBs (Dado et al.~2002, Dar and De 
R\'ujula~2004, Dado et al.~2009a,b), GRBs and their afterglows are 
produced 
by the interaction of bipolar jets of highly relativistic plasmoids (CBs) 
($\gamma\gg 1$) of ordinary matter with the radiation and matter along 
their trajectory 
(Shaviv and Dar~1995, Dar~1998). Such jetted CBs 
are presumably ejected in 
accretion episodes on the newly formed compact stellar object in 
core-collapse supernova (SN) explosions (Dar et al.~1992, Dar and 
Plaga~1999, Dar and De R\'ujula~2000), in merger of compact objects in 
close binary systems (Goodman et al.~1987, Shaviv and Dar~1995) and in 
phase transitions in compact stars (Dar~1998, Dar and De R\'ujula~2000, 
Dado et al.~2009b). For instance, in long GRBs produced in Type Ic 
supernova explosions it is hypothesized that an accretion disk or a torus 
is produced around the newly formed compact object, either by stellar 
material originally close to the surface of the imploding core and left 
behind by the explosion-generating outgoing shock, or by more distant 
stellar matter falling back after its passage (Dar and De R\'ujula~2000, 
2004). As observed in microquasars, each time part of the accretion disk 
falls abruptly onto the compact object, two CBs made of ordinary-matter 
plasma are emitted 
in opposite directions along the rotation axis from where matter has 
already fallen back onto the compact object due to lack of rotational 
support.  
The prompt $\gamma$-ray pulses and early-time X-ray flares are 
dominated by inverse Compton scattering (ICS) 
of glory photons - a light 
halo surrounding the progenitor star that was formed by stellar light 
scattered from the pre-supernova ejecta/wind blown from the progenitor 
star - by the CBs' electrons. The 
ICS is overtaken by synchrotron radiation (SR) when the CB enters the 
pre-supernova wind/ejecta of the progenitor star. The fast decline of the 
prompt emission with a fast spectral softening is overtaken (see, e.g., 
Dado et al.~2007, 2009a) by a broad-band SR afterglow with a much harder 
spectral index $\beta_X\simeq 1$, which stays constant in the X-ray band, 
when the CB propagates in the wind and interstellar (ISM) environment. ICS 
of the SR produces the emission of very high energy photons during the 
early time optical/NIR emission and the broad band afterglow (Dado and 
Dar~2009).

\section{Origin of correlations in the CB model} Detailed discussion of 
the origin of correlations between GRB properties can be found in Dar and 
De R\'ujula~2000, 2004, Dado et al.~2007, and Dado and Dar~2012a. In 
short, GRBs are not standard candles because of the diversity of their 
central engines and environments. But, because of the large bulk motion 
Lorentz factor $\gamma$ of the jet of CBs, their emitted radiation at 
redshift $z$, which is observed at a small angle $\theta$ relative to the 
direction of the jet, is boosted by a large Doppler factor 
$\delta\!=\!1/\gamma\,(1\!-\!\beta\,cos\theta)$ and collimated through 
relativistic beaming by a factor $\delta^2$. In the CB model, the large 
Doppler boosting, relativistic beaming and time aberration dominate the 
GRB properties and produce correlations between GRB observables, which 
depend on them (Dar and De R\`ujula~2000, 2004, Dado and Dar~2012a), 
despite their dependence on the CBs' intrinsic 
properties (rest frame properties indicated from now on by a prime) 
and on the environment along their 
trajectories (which produce a significant spread around these simple 
kinematic correlations).

\section{Correlations among and between  $\gamma$-ray and
         X-ray properties}
 
\subsection{Prompt emission}
\noindent 
{\bf Prompt $\gamma$-rays:} In the CB model, the prompt 
emission is a sum 
of individual pulses. For standard candle GRBs, in {\it each pulse} the 
peak energy $E_p$ of the time-integrated spectral energy flux,
the total gamma-ray energy emission under the assumption of isotropic 
emission $E_{iso}$, and  the peak luminosity $L_p$ satisfy 
$E'_p\propto \gamma_0\, \delta_0$, $E_{iso} \propto \gamma_0\,\delta_0^3$ 
and $L_p \propto \gamma_0^2 \delta_0^4$, respectively. 
Consequently, {\it each individual pulse in a GRB} is predicted to  
satisfy the triple correlation,  $L_p\propto E'_p\,E_{iso}$. The most 
probable viewing angle of GBs is $\theta\approx 1/\gamma_0$
yielding $\delta_0\approx \gamma_0$ and the approximate binary
power-law correlations $(1+z)\,E_p \propto [E_{iso}]^{0.5}$,
$(1+z)\,E_p \propto [L_p]^{0.33}$, and
$ L_p \propto [E_{iso}]^{1.5}$, respectively.
For XRFs that in the CB model are GRBs viewed far off-axis, i.e., 
$\theta\gg 1/\gamma$, the CB model yields  $(1+z)\,E_p \propto 
[E_{iso}]^{1/3}$,
$(1+z)\,E_p \propto [L_p]^{0.25}$, and $ L_p \propto [E_{iso}]^{4/3}$.   

For the
time-integrated $\gamma$-ray emission in multi-peak GRBs, 
the CB model
predicts for each pulse and for the entire GRB, roughly the same 
power-law  index $\approx 0.5$ for the $E'_p-E_{\gamma,iso}$ correlation.
The peak value of the luminosity integrated over a fixed observer 
time step, e.g., over 1 second, satisfies $L_{p,s}\propto  
\gamma_0\,\delta_0^3$ 
and consequently  $ E_{iso}\! \propto L_{p,s}$
and $(1+z)\,E_p\! \propto [L_{p,s}]^{0.5}$.
The above CB model correlations, which were predicted (Dar \& De R\'ujula
2000, 2004)
{\it before} their empirical discovery (Amati et al. 2002;
Yonetoku et al.~2004), are well satisfied by GRBs, as shown in Figs. 1-3.

\noindent
{\bf Prompt X-rays:}
In the CB model, the prompt 
emission is dominated by ICS of thin thermal 
bremsstrhalung (Dar \& De R\'ujula~2004) whose  time-integrated  
spectrum is given roughly by an exponentially cut-off power-law (CPL),
$\int E\, (dn/dE)\,dt \sim E^{1-\alpha}\,e^{\!-\!E/E_c}$, where
$\alpha\approx 1$, and then the
$E_p$, the peak energy of $E^2\,\int (dn/dE)\, dt$
is roughly the  "cut-off" energy $E_c$.
Consequently, for the  X-ray band, which 
is usually well below 
$E_p$, the emitted prompt X-ray energy, within a band width $\Delta E$
is given by  
\begin{equation}
E_{1,X}\approx {\Delta E \over E_p}\,  E_{\gamma,iso}\,,
\label{EX1}
\end{equation}
where $E_{\gamma,iso}=E_{iso}$ and $E_{1,X}$ are, respectively,  the 
prompt  gamma ray 
and X-ray energies
emitted under the assumption of isotropic emission.
Using  the CB model relations (Dar \& De R\'ujula~2000,2004),
$E'_p\propto \gamma_0\, \delta_0$ and 
$E_{iso} \propto \gamma_0\, \delta_0^3$, one obtains 
$E_{1,X} \propto  \delta_0^2.$ 

If the X-ray energy $E_{1,X}$ emitted during the first phase of the X-ray 
light-curve measured with Swift is proportional to the prompt X-ray 
emission, then 
Eq.~(1) implies that $E_{1,X} \propto [E_{\gamma, iso}]^{0.58\pm 0.08}$, 
where the exact power-law index depends on the mixure of ordinary GRBs and 
XRFs in the observed sample. This predicted index is in agreement with the 
correlation $E_{1,X} \propto [E_{iso}]^{0.56\pm 0.04}$ reported in 
Margutti et al.~2013.

\subsection{The X-ray afterglow}   
In the CB model, the X-ray  afterglow is 
dominated by synchrotron radiation (SR), and
begins when the (merged) CBs enter the circumburst 
wind, which have been blown by the progenitor star sometime before the SN
explosion. 
The spectral energy density of the {\it unabsorbed} 
X-ray afterglow of a single CB has the form (see, e.g., Eq.~(26) in Dado 
et al.~2009a),
\begin{equation}
F_{\nu} \propto  n^{(\beta_X+1)/2}\,
[\gamma(t)]^{3\,\beta_X-1}\, [\delta(t)]^{\beta_X+3}\, \nu^{-\beta_X}\, ,
\label{Fnu}
\end{equation}
where $n$ is the baryon density of the medium, 
and $\beta_X+1=\Gamma_X$ is 
the photon spectral index of the emitted (unabsorbed) radiation.
For $\gamma^2 \gg 1$ and $\theta^2
\ll 1$, $\delta \approx 2\, \gamma/(1\!+\!\gamma^2\, \theta^2)$ to an
excellent approximation. 
The X-ray band is well above the break frequency, where typically 
$\Gamma_X\approx 2$, i.e., $\beta_X \simeq 1$, and Eq.~(2) yields
$F_{\nu} \propto  n\,[\gamma]^2\, [\delta]^4\, \nu^{-1}$.
For a "shot-gun" configuration of CBs, Eq.~(2) yields (Dado and Dar 
2012b) 
\begin{equation}
F_{\nu} \propto n^{(\beta_X+1)/2}\,[\gamma(t)]^{4\,\beta_X}\, 
\nu^{-\beta_X}\,, 
\label{Fnusg}
\end{equation}
which reduces to 
$F_{\nu} \propto  n\,[\gamma(t)]^4\,\nu^{-1}$ for $\beta_X\simeq 1$.

\subsection{The early-time and break-time luminosities of canonical AGs}

The intercepted ISM particles that are swept into the CB decelerate its 
motion. For 
a CB of a baryon number $N_{_B}$, a radius $R$ and an initial Lorentz 
factor $\gamma_0=\gamma(0)\gg 1$, which propagates in an ISM of {\it a 
constant 
density} $n$, relativistic energy-momentum conservation 
yields the deceleration law  (Dado et al. 2009b and references therein)
\begin{equation}
\gamma(t) = {\gamma_0\over [\sqrt{(1+\theta^2\,\gamma_0^2)^2 +t/t_0}
          - \theta^2\,\gamma_0^2]^{1/2}}\,,
\end{equation}
where $t_0={(1\!+\!z)\, N_{_{\rm B}}/ 8\,c\, n\,\pi\, R^2\,\gamma_0^3}\,.$

As long as  $t \lsim t_b\!=\!(1\!+\!\gamma_0^2\theta^2)^2\,t_0$,
$\gamma(t)$ and $\delta(t)$  
change rather slowly with $t$, which generates the plateau phase of 
$F_\nu(t)$ of canonical X-ray AGs that was predicted by the CB model
(see, e.g., Dado et al.~2002, Figs.~27-33) and later observed with 
Swift (Nousek et al.~2006, Panaitescu et al.~2006, Zhang et al.~2006).   
For $\beta_X\simeq 1$, the X-ray  luminosity at 
the beginning time $t_i$ of the plateau 
phase that follows from Eq.~(2) is given by 
$L_X(t_i)\propto \gamma_0^2\, \delta_0^4$.
From Eq.~(4) it follows that for $t\gg$  $t_b$,  
$\gamma(t)\!\rightarrow\!\gamma_0 (t/t_b)^{-1/4}$,
$[\gamma(t)\theta]^2$ becomes $\ll\! 1$ and  $\delta\!\approx\! 
2\,\gamma(t)$, which result in a late-time power-law decay 
\begin{equation}
F_\nu(t)\propto [\gamma_0]^{4\,\beta_X+2}\, (t/t_b)^{-\beta_X-1/2}\,.
\label{latefnu}
\end{equation}
Hence,  the  dependence on the L\&D factors of the 
break-time is 
$t'_b\propto 1/\gamma_0\, \delta_0^2$, which yields the triple 
correlation 
\begin{equation}
t'_b\propto 1/[E'_{p}\, E_{iso}]^{1/2}.
\label{tbepeiso} 
\end{equation}
Then, the substitution $t_b=(1+z)\,t'_b $ and $E'_p \propto 
[E_{iso}]^{1/2}$
yields the approximate pair  correlations
\begin{equation}
t_b/(1+z) \propto [E_{iso}]^{-0.75}\propto [E'_p]^{-1.5} \, .
\label{tbeiso} 
\end{equation}
Note that the predicted late-time ($t'\gg t'_b$) behaviour of the 
X-ray luminosity at a fixed $t'$,  
\begin{equation}
L(t')\propto \gamma_0^{1.5}\, t'^{-1.5}\propto 
[E_{iso}]^{0.5}\,t'^{-1.5}\,,
\label{lateL}
\end{equation}
is in agreement with the observed behaviour 
$L_X(11h)\propto [E_{iso}]^{0.50}$  (Margutti et al.~2013).

Note also that for a 'shotgun' configuration of CBs,
$L_X(t_i)\propto \gamma_0^{4\, \beta_X}$ and 
$L_X(t\gg t_b) \propto \gamma_0^{4\,\beta_X}\, (t/t_b)^{-\beta_X}$ (Dado 
\& Dar 2012b), and
the late-time power-law behaviour of $L_X$  extrapolated back to
$t=t_b$ yields $L_X(t_b) \propto \gamma_0^{4\,\beta_X}$.
Hence, for $\beta \simeq 1$,  
roughly $L_X(t_b)\propto L_X(t_i)\propto \gamma_0^{5\pm 1}$.

The CB model (achromatic) break-time parameter $t_b$, however, is not the 
same as the chromatic break-time parameter of the heuristic smooth 
broken power-law function used by Margutti et al.~(2013) to 
parametrize canonical AGs. Consequently, the correlations satisfied by the 
CB model break-time of the X-ray AG may differ slightly from those 
inferred from the phenomenological broken power-law fits reported in 
Margutti et al.~2013.

\subsection{$E_{2,X}$ in canonical GRBs} 
Assuming isotropic emission in the GRB rest frame
at redshift $z$ (luminosity distance $d_L$),
the total emitted energy during the afterglow 
phase in the rest frame X-ray band 
$[\nu'_1,\nu'_2]$ is given by
\begin{equation}
E_{2,X}={4\, \pi\ d_L^2\over(1+z)}\, \int \int F_\nu(t)\, dt\, d\nu 
\label{Exiso}
\end{equation}
where the $\nu$ integration is from 
$\nu_1=\nu'_1/(1+z)$ to  $\nu_2=\nu'_2/(1+z)]$. Since
$F_\nu \propto \nu^{-\beta_X}\sim \nu^{-1}$, the $\nu$ integration is 
practically independent of redshift. Moreover,
Eq.~(4) can be used to convert the 
time integration  in Eq.~(9) to integration over $\gamma$, yielding
$E_{2,X} \propto \gamma_0^2\, \delta_0$ for the total energy 
emitted in the 0.3-30 keV X-ray band during the afterglow phase, 
assuming isotropic emission. 

\section{Comparison with observations}
\subsection{Binary correlations}
Table~1 Summarizes the dependence in the CB model of $\gamma$-ray and 
X-ray properties 
of GRBs on the bulk motion L\&D factors of the CBs, 
which was detailed in section 4.
In Table~1, we have used the notation $E'_p$ and $L_{iso}$ 
for the rest frame peak 
gamma ray energy and the mean luminosity during T90, respectively, of 
the prompt $\gamma$-ray emission. 

Table~2 presents a  comparison between the best fit indices of the 
21 observed power-law correlations among seven chosen observables
and those predicted by  the  cannonball model. The predicted indices 
are the arithmetic mean of the typical cases  
$\theta\,\gamma_0\approx 1$ and $\theta^2\, \gamma_0^2\gg 1.$

Table~3 summarizes  the observed correlations between  $E'_p$, 
$E_{iso}$  and $t_b$ and the power-law correlation 
indices expected in the cannonball model and in the collimated 
fireball model. 

Figs. 1-3 present our best fitted power-laws for the
$E'_p-E_{iso}$,  $E'_p-L_{iso}$, and
$L_{iso}- E_{iso}$ correlations for a
sample of 96 GRBs  with known redshift, which  was
compiled by  Yonetoku  et al. (2010).
Using essentially the method advocated by  D'Agostini (2005),
we obtained the best fit correlations
$(1+z)E_p \propto  [E_{iso}]^{0.529}$,
$(1+z)E_p  \propto [L_{iso}]^{0.532}$, and
$L_{p,\gamma}  \propto [E_{iso}]^{1.13}$  in good agreement with 
those
predicted by the CB model.  Note that the correlations
satisfied by $E_{p} $ imply that $ E_{iso}\propto 
[L_{iso}]^{1.01}$.
The $\sim 10\%$  difference in the power-law index of the 
$E_{iso}-L_{iso}$  correlation probably provides 
a  realistic estimate of the accuracy of the power-law indices  extracted 
from the observational data.  

Fig.~4 compares the triple 
$t'_b-E'_p-E_{iso}$ correlation predicted by the CB model (Eq.~6) and the 
observed
correlation in  68 Swift GRBs (Evans et al.~2009) from the above GRB
sample, which have  a good Swift/XRT temporal coverage of their X-ray
afterglow
during the first day (or more) following  the prompt emission phase
and  have no superimposed flares. In this sample, the X-ray afterglow
of 54 GRBs  clearly shows a break and  no  afterglow-break was observed in
14 GRBs.
In order not to bias the values of $t_b$, $E_{p}$, and $E_{iso}$ by
the CB model fits, the break times were taken to be the times of the first
break with $\alpha(t<t_b)<\alpha(t>t_b)$ obtained from the broken
power-law fit to the GRB X-ray afterwglow measured with the Swift/XRT 
and reported in the Leicester XRT GRB Catalogue (Evans et al.~2009).
The Spearman rank (correlation coefficient)  of the  
$t'_b - (E'_p\,E_{iso})$ correlation is $r= -0.65$
with a chance probabilities less than  $2.6\times 10^{-9}$.
The best fit triple correlation
$t'_b \propto 1/[E'_p\, E_{iso}]^p $ yields $p=-0.58 $

The approximate binary correlations $t'_b-E_{iso}$ and $t'_b-E'_p$ 
that were
obtained by substitution of the CB model predicted correlation $E'_p
\propto [E_{iso}]^{1/2}$ in the triple correlation $t'_b-E'_p-E_{iso}$
(Eq.~6), are compared with the observational data in Figs.~5 and 6,
respectively. GRB 980425 was excluded from the GRB sample because it is an
outlier with respect to the binary $E'_p-E_{iso}$ correlation (as expected
in the CB model). The Spearman ranks of the observed $t_b/(1+z)-E_{iso}$ 
and $t_b/(1+z)-(1+z)E_p$ correlations are -0.49 and -0.63 with chance
probabilities less than $4.5\times 10^{-4}$ and $1.0\times 10^{ -6}$,
respectively. As expected
in the CB model, they are larger than that of the
$t'_b-(E'_p\, E_{iso})$ correlation.
The best fit power-law indices of the $t'_b-E_{iso}$ and $t'_b-E'_p$
correlations  are $p=-0.70\pm 0.06$  and $p=-1.64 \pm 0.04$, 
respectively,
consistent with their  predicted values by the CB model,
$-0.75$ and $-1.50$, respectively.

The best fit power-law indices $p=0.54\pm 0.01$ , $p=-1.61\pm 0.04$, and
$p= -0.70\pm 0.06$ of the observed $E'_p-E_{iso}$, $t'_b-E'_p$ and
$t'_b-E_{iso}$ power-law correlations, respectively, are at odds
with the values 1, -1, and -1, respectively, expected in the
conical fireball model.

Table~2 compares the approximate power-law correlations between the 
$\gamma$-ray and X-ray propreties of long GRBs that are expected in the CB 
model (upper rows) from their dependence on the L\&D factors and those 
extracted in the limit of small dispersions from the 
correlations reported in Margutti et al.~2013 (lower rows).
As can be seen from Table~2, 
the correlations predicted by the CB model agree well with those 
inferred by Margutti for LGRBs.

\subsection{The triple correlation $E_{X,iso}-E_{\gamma,iso}-E'_p$}
A triple correlation $E_{X,iso} \propto [E_{\gamma,iso}]^{1.06 \pm 
0.06}/[E'_p]^{0.74 \pm 0.10}$ was found  by Bernardini et al.~(2012)
to be well satisfied  by both long and short GRBs. This correlation was 
updated in  Margutti et al.~(2013) to be 
$E_{X,iso} \propto [E_{\gamma,iso}]^{1.00 \pm  
0.06}/[E'_p]^{0.60 \pm 0.10}$. 
In the CB model, such triple correlations that unite
LGRBs and SHBs are a simple combination  of corresponding 
binary power-law correlations of kinematic origin 
satisfied by  LGRBs and SHBs (with the same index but different 
normalization). In particular, the above triple correlation 
is a simple consequence of the correlation $E'_p\propto 
[E_{\gamma,iso}]^p$  (shown in Fig.~1 for LGRBs 
with the current best fit value $p=0.54$)
that was predicted by the CB model to be satisfied  
both by LGRBs  (e.g., Dar and De R\'ujula 2000,2004)
and by SHBs (Dado et al.~2009b, Fig.~5) and the   
pair  correlation $E_{X,iso}\propto [E_{\gamma,iso}]^{0.67\pm
0.01}$ that was discovered by Margutti et al.~(2013) for LGRBs.
To see that, let us rewrite the 
$E_{X,iso}-E_{\gamma,iso}$ correlation that was found by 
Margutti et al.~(2013) for LGRBs in the form 
\begin{equation}
E_{X,iso}\propto [E_{\gamma\,iso}]^{0.67}\, {[E_{\gamma\,iso}]^m\over
                  [E'_p]^{m/p}}\,,
\end{equation}
where the second factor on the right hand side (RHS)
is a constant. 
For the triple correlation reported by    
Margutti et al.~(2013), their best fit implies
$m=0.33\pm 0.06$ and $m/p=0.60\pm 0.10$, which yield
$p=(0.33\pm 0.06)/(0.60 \pm 0.10)=0.55\pm 0.07$.
The best fit values reported by Bernardini et al.~(2012)  
yield
$m=0.39\pm 0.06$ and $m/p=0.74\pm 0.10$, which implies
$p=(0.39\pm 0.06)/(0.74 \pm 0.10)=0.527\pm 0.05$.  Both values are 
consistent with the CB model prediction $p\approx 0.5$ and with
$p=0.526$ obtained from the best fit 
shown in Fig.~1. For LGRBs,  the triple correlation as written in  
Eq.~(10) is independent of the choice of an $m$ value. However, in the CB 
model the power-law indices of  kinematic  correlations  are common  
to  SHBs and LGRBs.
Thus, the value of $m$ can be adjusted   such that 
the triple correlation in LGRBs is satisfied also by SHBs.
       
\subsection{The triple correlation $t'_b-E'_p-E_{\gamma,iso}$}

As explained in section 3, the strong dependence of GRB observables on 
both the Lorentz factor and the Doppler factor 
of the jetted CBs yields triple-correlations among these observables,  
which  can be reduced to  binary correlations only with additional
assumptions. We demonstrate this for an important case
- the triple correlation $t'_b-E'_p-E_{\gamma,iso}$ 
as summarized in Eq.~(6), which is based on the decelaration origin of 
the break in the canonical afterglow of the CB  model. 
In Fig.~4, this correlation is compared  with observations of 68 Swift 
GRBs/XRFs with known redshift, good temporal coverage of their X-ray 
afterglow  during the first day (or more) following the prompt emission 
phase, and  well measured  $E_p$ and $E_{iso}$
with  Konus-WIND  and/or Fermi GBM. 
In order not to bias the values of $t'_b$, $E'_{p}$  and 
$E_{iso}$ by CB model fits, 
the break times were taken to be the times of the 
first break with $\alpha(t<t_b)<\alpha(t>t_b)$ 
obtained from the broken power-law fits to the GRB 
X-ray afterglow measured with the Swift/XRT
and reported in the 
Leicester XRT  GRB Catalogue (Evans et al.~2009).
The values of $E_{p,\gamma}$ and $E_{\gamma,iso}$ were
adopted from  communications of the 
Konus-Wind and Fermi GBM collaborations 
to the  GCN Circulars Archive (Barthelmy~1997), and from 
publications by Amati et al.~(2007, 2008 ), Yonetoku et al.~(2010),
Gruber et al.~(2011),  Nava et al.~(2012), and D'Avanzo et al.~(2012).
As shown in Fig.~4, the triple correlations predicted by the CB model
is well satisfied by the observational data;
the best fit power-law  $t'_b\propto [E'_p\,E_{\gamma,iso}]^p$  
yields $p=-0.58\pm 0.04$ in good agreement with the 
predicted power-law index  $p=-1/2$.

The binary correlations $t_b/(1+z)\propto [E_{iso}]^{-0.75}\propto 
[E'_p]^{-1.5}$  that were obtained by substituting the CB model 
correlation 
$E'_p\propto E_{iso}^{1/2}$ in the triple 
correlation (Eq.~6) are in good agrement with the 
the best fit power-law correlations  
shown in Figs.~5 and 6, with power-law 
indices $-0.69\pm 0.06$ and $-1.62\pm 0.04$, respectively.
 
In particular, the break-time correlations imply that GRBs with very large 
values of
$E_{iso}$  and $E'_p$ have a small $t'_b$ value,  which, probably, 
is hidden under the tail of the prompt emission or precedes the start of 
the XRT observations (Dado et al.~2007). Indeed, the X-ray afterglow of 
all the GRBs in our sample, which have  large values 
of $E_{iso}$ and $E'_p$, such as 061007, 080319B and 130427A,
have  a power-law decline consistent with the post break power-law decline 
predicted by the CB model (see, e.g., Dado et al. 2007, 2009a).  For such 
GRBs, the observations provide only upper bounds on the break time of 
their X-ray afterglow, which  are indicated by down arrows in 
Figs.~4-6.

\section{Discussion}
  
A major breakthrough in the study of gamma ray bursts (GRBs) was the 
discovery of their X-ray afterglows by the Beppo-SAX satellite (Costa et 
al.~1997) that allowed their arcminute localization and consequently the 
discovery of their longer wave-length afterglows (van Paradijs et 
al.~1997, Frail and Kulkarni~1997), which were predicted (Paczynski and 
Roads 1993, Katz~1994, Meszaros and Rees~1997) by the fireball model (FB) 
of GRBs (Paczynski~1986, Goodman~1986). Consequently, the fireball model 
was widely accepted as the correct model of GRBs and their afterglows 
(e.g., Meszaros~2002, Zhang and Meszaros~2004, Zhang~2007). The rich data 
on GRBs and their afterglows obtained in recent years with the Swift and 
Fermi satellites, complemented by data from ground-based rapid response 
telescopes and large follow-up telescopes, have, however, challenged 
this 
prevailing view (e.g., Covino et al.~2006, Curran et al.~2006, Burrows and 
Racusin~2006, Kumar et al.~2007, Zhang et al.~2007, Liang et al.~2008, 
Godet \& Mochkovitch~2011, Margutti et al.~2013, and references therein).

In contrast, the cannonball model (CB) of GRBs has been very successful in 
predicting the general properties of the prompt emission and afterglows of 
both long and short GRBs and in reproducing their detailed light-curves 
(e.g., Dado et al.~2009a,b and references therein).  This success was 
despite the apparently different origins and environments of long GRBs and 
SHBs. But, it involved an adjustment of free parameters for each GRB, 
which could have made one wonder whether the agreement between theory and 
observations was due to the flexibility of the model rather than its 
validity. Many properties of GRBs, correlations among them, and closure 
relations that are predicted by the CB model, however, do not involve 
adjustable parameters and thus enable more stringent tests of the model. 
So far the CB model was able to predict correctly all the main established 
correlations among GRB observables (see, e.g., Dar and De 
R\'ujula~2000,2004, Dado and Dar~2012a,b and references therein) including 
the newly discovered correlations (Liang et al.~2010) among and between 
the prompt $\gamma$-ray and optical properties of GRBs (e.g., Dado and 
Dar~2012a and references therein). In this paper, we have shown that 
the correlations that were discovered 
recently between the $\gamma$-ray and X-ray properties of GRBs (Margutti 
et al.~2013, Bernardini et al. 2012) are correctly predicted by the CB 
model. Moreover, we have also shown that 
the triple correlation $E_{X,iso}-E_{\gamma,iso}-E'_p$ that is satisfied 
by both LGRBs and SHBs (Bernardini et al.~2012, Margutti et al.~2013) 
probably is a simple consequence of the fact that the binary power-law 
correlations $E_{X,iso}-E_{\gamma,iso}$ and $E'_p-E_{\gamma,iso}$ are 
satisfied by both LGRBs and SHBs with roughly the same power-law index and 
different normalizations, as expected in the CB model.
 
Finally, we note that, in contrast to the cannonball model, the conical 
fireball model has not been able to explain  the observed 
correlations between the prompt and afterglow emissions, in particular 
those which involve  the afterglow break-time, as was shown in 
detail in this paper.

{\bf Acknowledgment}: We thank Raffaela Margutti and an anonymous 
referee for useful comments, and David Gruber for useful 
communications.

\begin{deluxetable}{cllllllll}   
\tablewidth{0pt}
\tablecaption{The dependence on Lorentz and Doppler factors of 
LGRB properties in the CB model}  
\tablehead{
\colhead{property:} & \colhead{$E_{1,X}$} & \colhead{$E_{2,X}$} 
&\colhead{$L_X(t_i)$} & \colhead{$L_X(t'_b)$} & \colhead{$t'_b$} &
\colhead{$E_{\gamma,iso}$} & \colhead{$E'_{p}$} & \colhead{$L_{p,s}>$}} 

\startdata
propto: & $\delta_0^2$ & $\gamma_0^2\,\delta_0$ 
&$\gamma_0^2\,\delta_0^4$& $\gamma_0^3\,\delta_0^3$ 
& $1/\gamma_0\,\delta_0^2$ & $\gamma_0\, \delta_0^3$& 
$\gamma_0\,\delta_0$ & $\gamma_0\, \delta_0^3$\\
\hline
\enddata
\end{deluxetable}

\begin{deluxetable}{clllllll}
\tablewidth{0pt}
\tablecaption{Comparison between the power-law index $m$ of the 
correlations $Y\propto X^{m}$  predicted by the CB model (upper rows) 
for various pairs ($X,Y$) of properties of long duration GRBs 
and those obtained  (lower rows)  
by Margutti et al. (2013) from a comprehensive analysis of the X-ray light
curves of more than 650 GRBs measured with the Swift X-ray telescope
(XRT) during the years 2004-2010. The correlations 
satisfied by the afterglow break-time  were inferred from the Swift/XRT 
light-curves (Evans et al.~2009) by the authors of the present paper}. 
\tablehead{
\colhead{Y} & \colhead{$E_{2,X}$} & \colhead{$L_X(t'_i)$} & 
\colhead{$L_X(t'_b)$} & \colhead{$t'_b$} & 
\colhead{$E_{iso}$}&
\colhead{$E'_p$} & \colhead{$L_{p,s}$}}
\startdata
\hline
$E_{2,X} $ & 1& 0.50 & 0.50& -1& 0.71& 1.50 & 0.60\\
           &1 & 0.46& 0.50&-0.99& 0.74& 1.48& 0.60 \\
\hline          
$L_X(t'_i)$ & 2    & 1 & 1     & -2.0 &  1.5 & 3    & 1.5 \\
            & 2.17 & 1 & 1.09 &-1.52&  1.52 & 2.71 & 1.31\\
\hline
$L_X(t'_b)$ & 2   & 1    & 1 & -2.0  & 1.50 &  3   & 1.00 \\
            &  2  & 0.92 & 1 & -1.71 & 1.06 & 2.02 &1.06\\
\hline
$t'_{b}$  &  -1. & -0.50& -0.50 & 1& -0.75& -1.50 & -0.75\\ 
          &-1.07 & -0.66& -0.63& 1&  -0.70& -1.61& -0.61\\
\hline
$E_{iso}$  &1.41& 0.67&  0.67 &-1.33 & 1 & 2 & 1 \\
               & 1.35& 0.66& 0.63  &-1.43 & 1 & 1.90 & 0.88 \\
\hline
$E'_p$& 0.67 & 0.33 &0.33& -0.67& 0.50  & 1 & 0.50 \\
        & 0.78& 0.37& 0.36& -0.62 & 0.53 &1& 0.52 \\
\hline
$L_{p,s}$ & 1.67 & 0.67 & 1.00  &-1.33 &   1.& 2.0  & 1 \\
      & 1.59 & 0.76 & 0.86 & -1.43 &   1.13&  2.14 & 1 \\
\hline
\enddata
\end{deluxetable}

\begin{deluxetable}{cllllll}   
\tablewidth{0pt}
\tablecaption{Summary  of the observed power-law correlations 
between  $E'_p$, $E_{iso}$, and $t'_b$ and their power-law indices  
expected in the CB model and in the collimated FB model.
$\rho$ is the Spearman rank (corrrelation coefficient), $P(\rho)$ is the 
probability to obtain by chance  a correlation coefficient larger than  
$|\rho|$, and $p$ is the power-law index of the correlation.}
\tablehead{\colhead{Correlation} & \colhead{$\rho$}&
\colhead{$P(\rho)$} & \colhead{$p(obs)$}&\colhead{$p(CB)$}&\colhead{$p(FB)$}} 
\startdata
\hline
$E'_p-E_{iso}$ & +0.87& $\approx 0$ & $+0.54\pm 0.01$ & +0.50 & +1 \\

$t'_b-(E'_p\, E_{iso})$ & -0.70& $\approx 2.6\times 10^{-9}$ & 
$-0.58\pm 0.04$ & -0.50 &     \\

$t'_b-E'_p$ & -0.63& $\approx 1.0 \times 10^{-6}$ & 
$-1.62\pm 0.04$ & -1.50 & -1.0   \\

$t'_b-E_{iso}$ & -0.49& $\approx 4.5\times 10^{-4}$ 
& $-0.69\pm 0.06$ & -0.750 &  -1.0 \\
\hline
\enddata
\end{deluxetable}

\newpage
\begin{figure}[]
\centering
\vspace{-2cm}
\epsfig{file=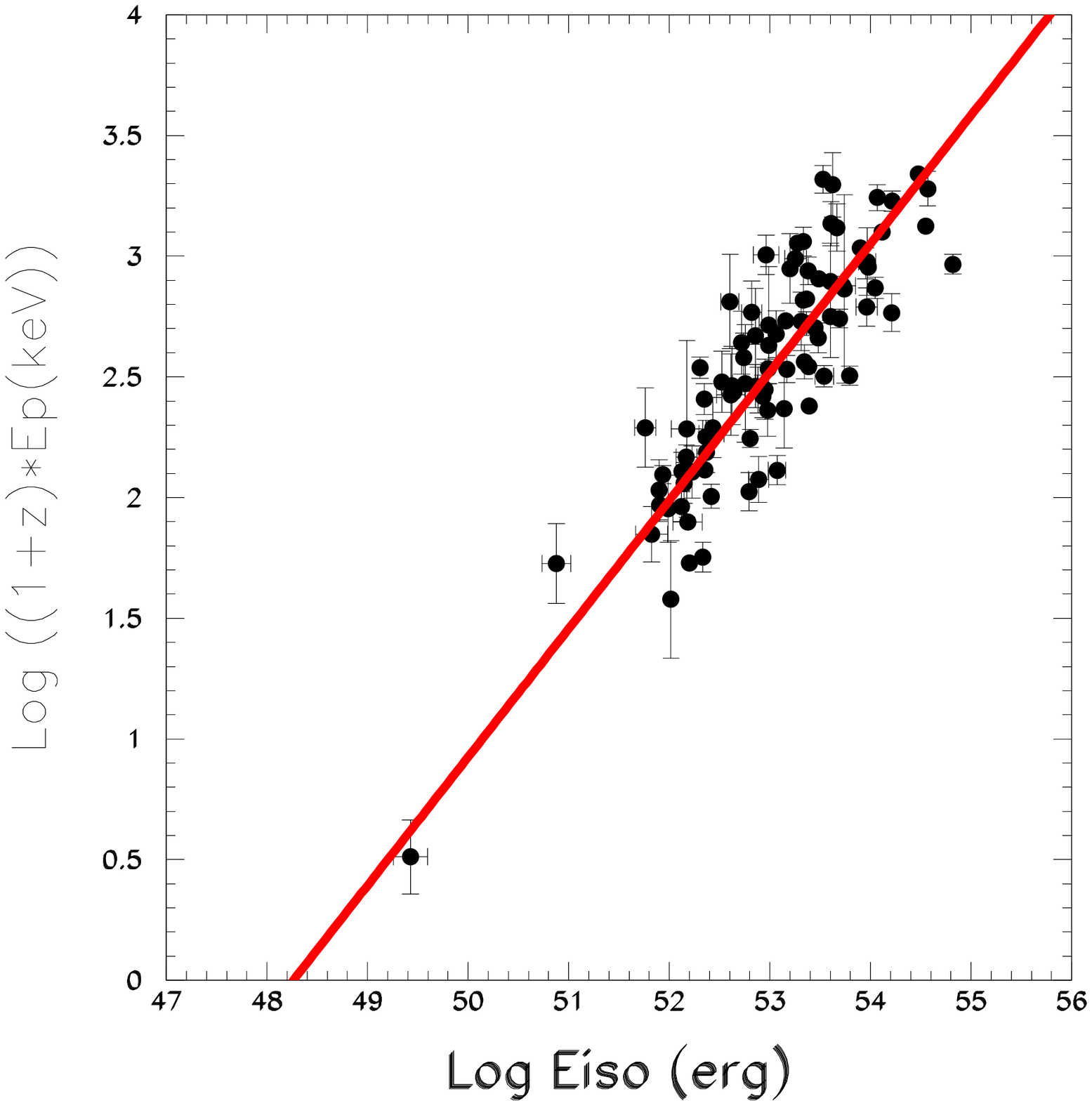,width=16.cm,height=16.cm}
\caption{
The observed correlation between $(1+z)\, E_p$ and 
$E_{iso}$
for 121 GRBs with known redshift.
The best fit power-law correlation (straight line) has a power-law index 
0.54.}
\label{FIGC1}
\end{figure}

\newpage
\begin{figure}[]
\centering
\vspace{-2cm}
\epsfig{file=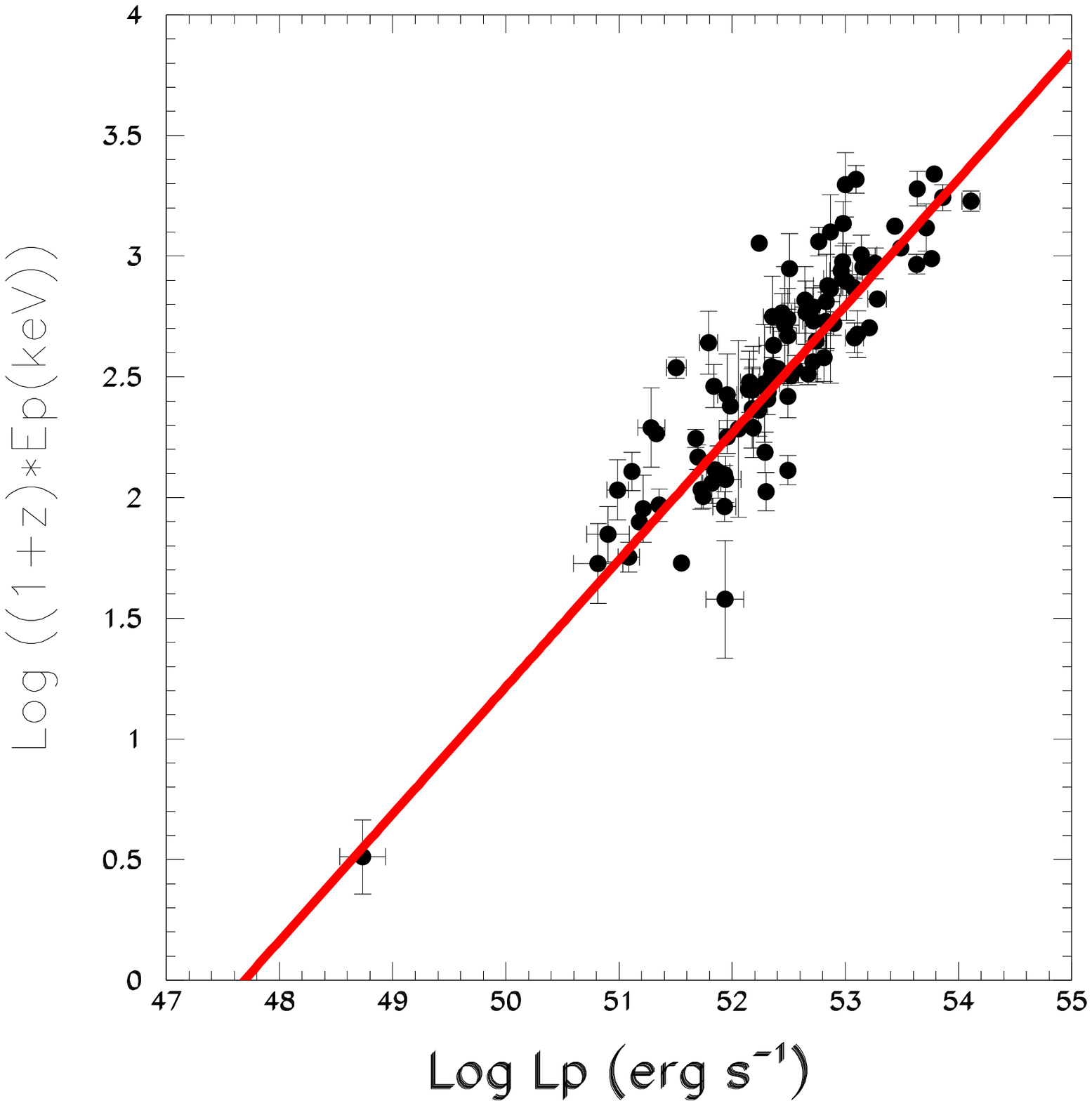,width=16.cm,height=16.cm}
\caption{
The observed correlation between $(1+z)\, E_p$ and $L_{p,s}$
for 121 GRBs with known redshift. The best fit power-law correlation 
(straight line) has a power-law index 0.526.}
\label{FIGC2}
\end{figure}

\newpage
\begin{figure}[]
\centering
\vspace{-2cm}
\epsfig{file=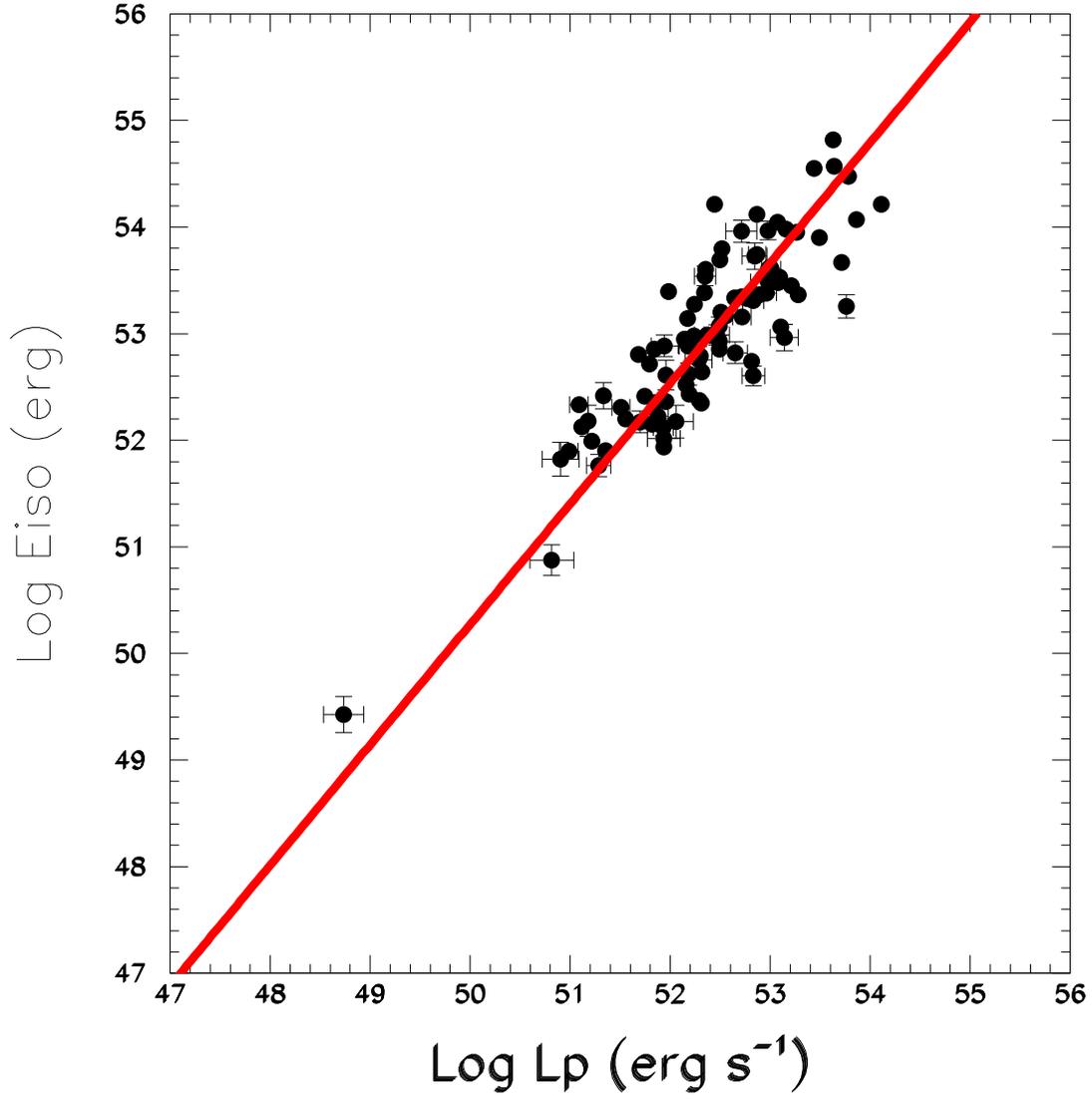,width=16.cm,height=16.cm}
\caption{
The observed correlation between 
$E_{\gamma,iso}$ and $L_{p,s}$ 
for 121 GRBs with known redshift.
The best fit power-law correlation (straight line) has  a power-law index 
1.13.}
\label{FIGC3}
\end{figure}

\newpage
\begin{figure}[]
\centering
\vspace{-2cm}
\epsfig{file=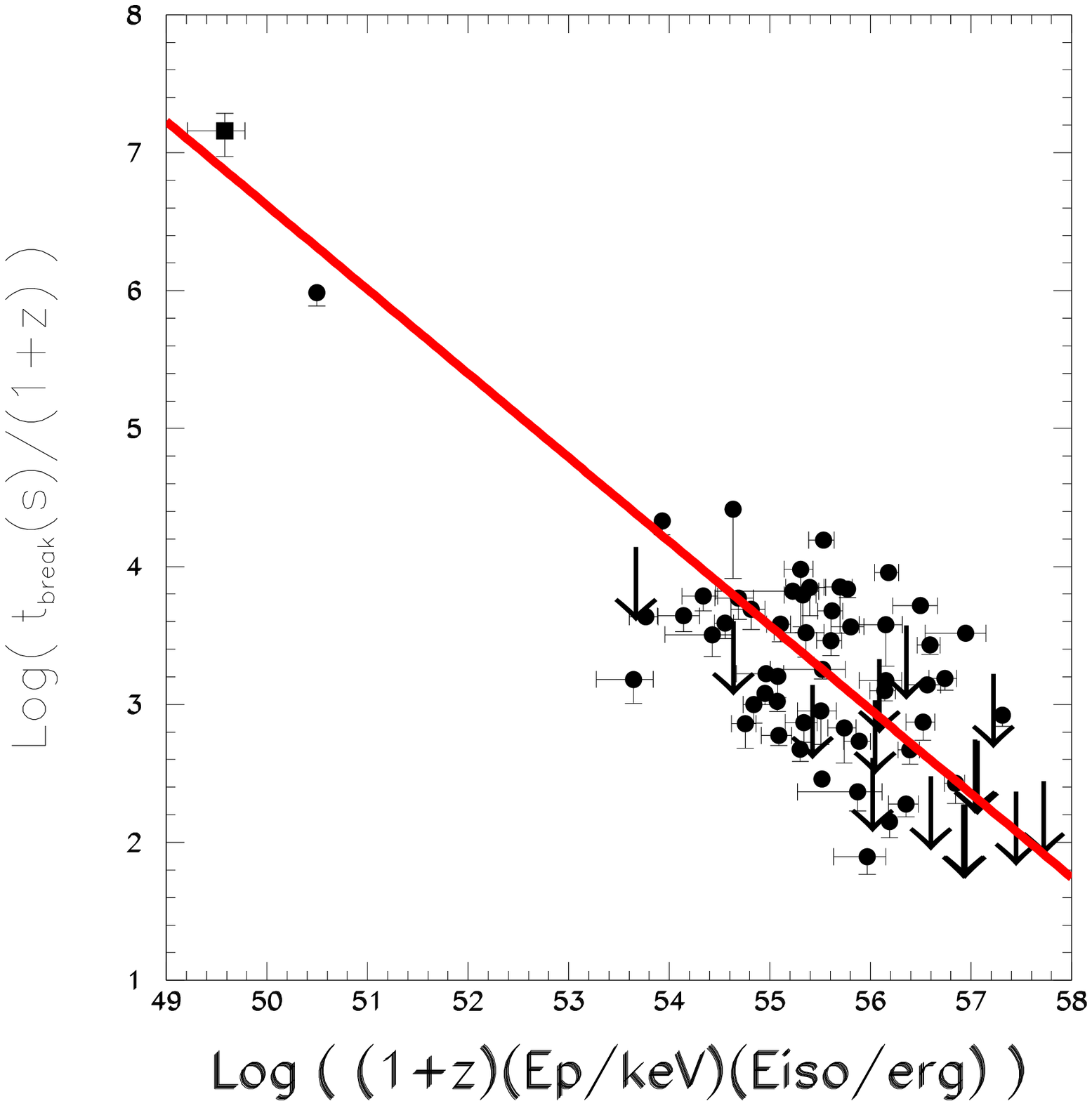,width=16.cm,height=16.cm}
\caption{The observed triple correlations 
$t'_b-E'_p-E_{iso}$ 
in 68 Swift GRBs with measured  redshift, 
$t'_b$,  $E'_p$, and $E_{iso}$ and its best fit power-law
(straight line with a power-law index -0.58). Arrows 
indicate observational upper bounds on early-time deceleration breaks
before the beginning of the Swift/XRT observations or hidden under the 
prompt emission tail. }
\label{FIGC4}
\end{figure}

\newpage
\begin{figure}[]
\centering
\vspace{-2cm}
\epsfig{file=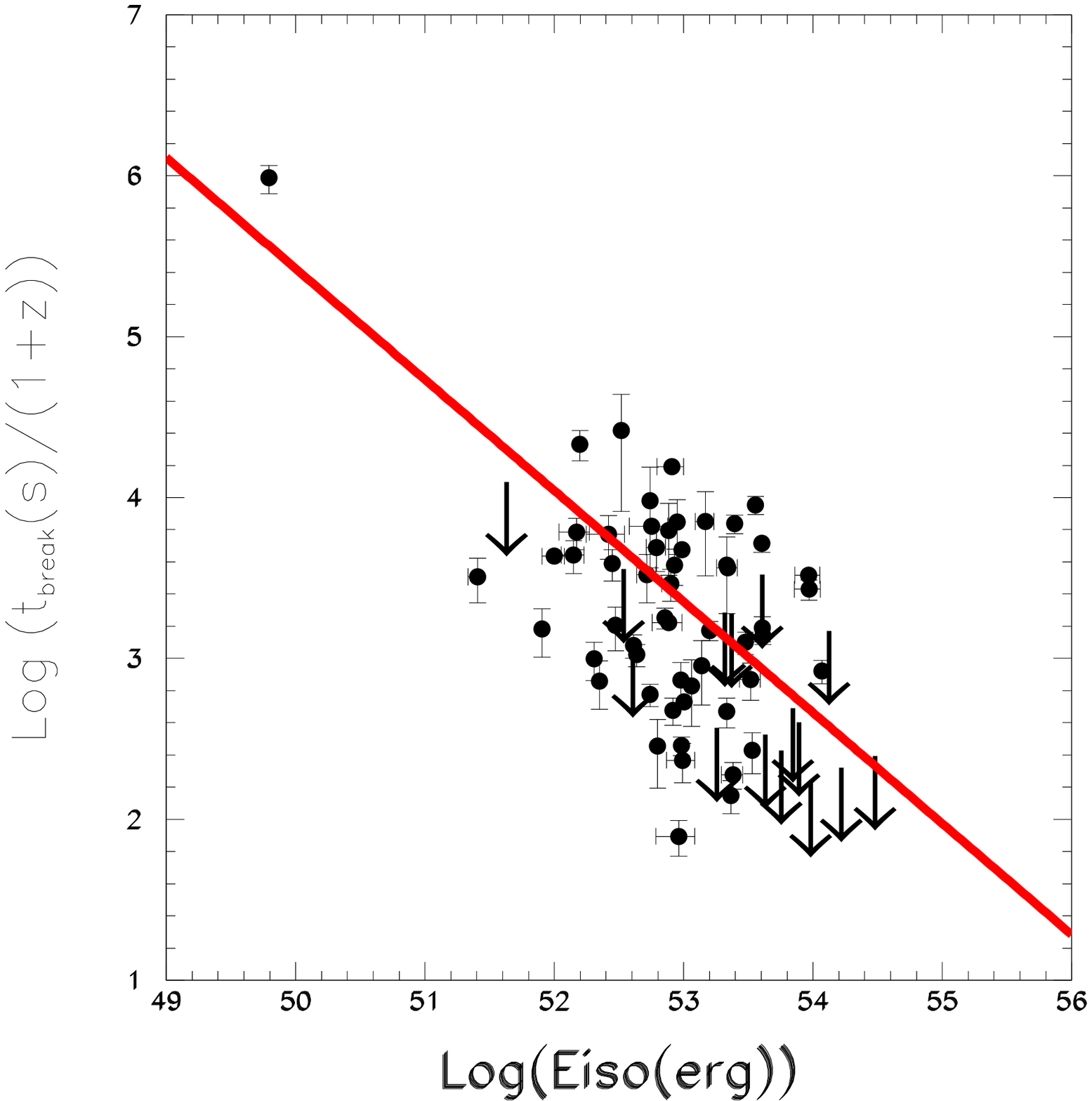,width=16.cm,height=16.cm}
\caption{The binary correlation           
$t_b/(1+z)-E_{iso}$  observed in
67 Swift GRBs with measured  redshift, 
$t'_b$ and $E_{iso}$ and its best fit 
power-law (straight line with a power-law index -0.70). 
Arrows indicate 
observational upper bounds on early time deceleration breaks
before the beginning of the Swift/XRT observations or hidden under the 
prompt emission tail.}
\label{FIGC5}
\end{figure}

\newpage
\begin{figure}[]
\centering
\vspace{-2cm}
\epsfig{file=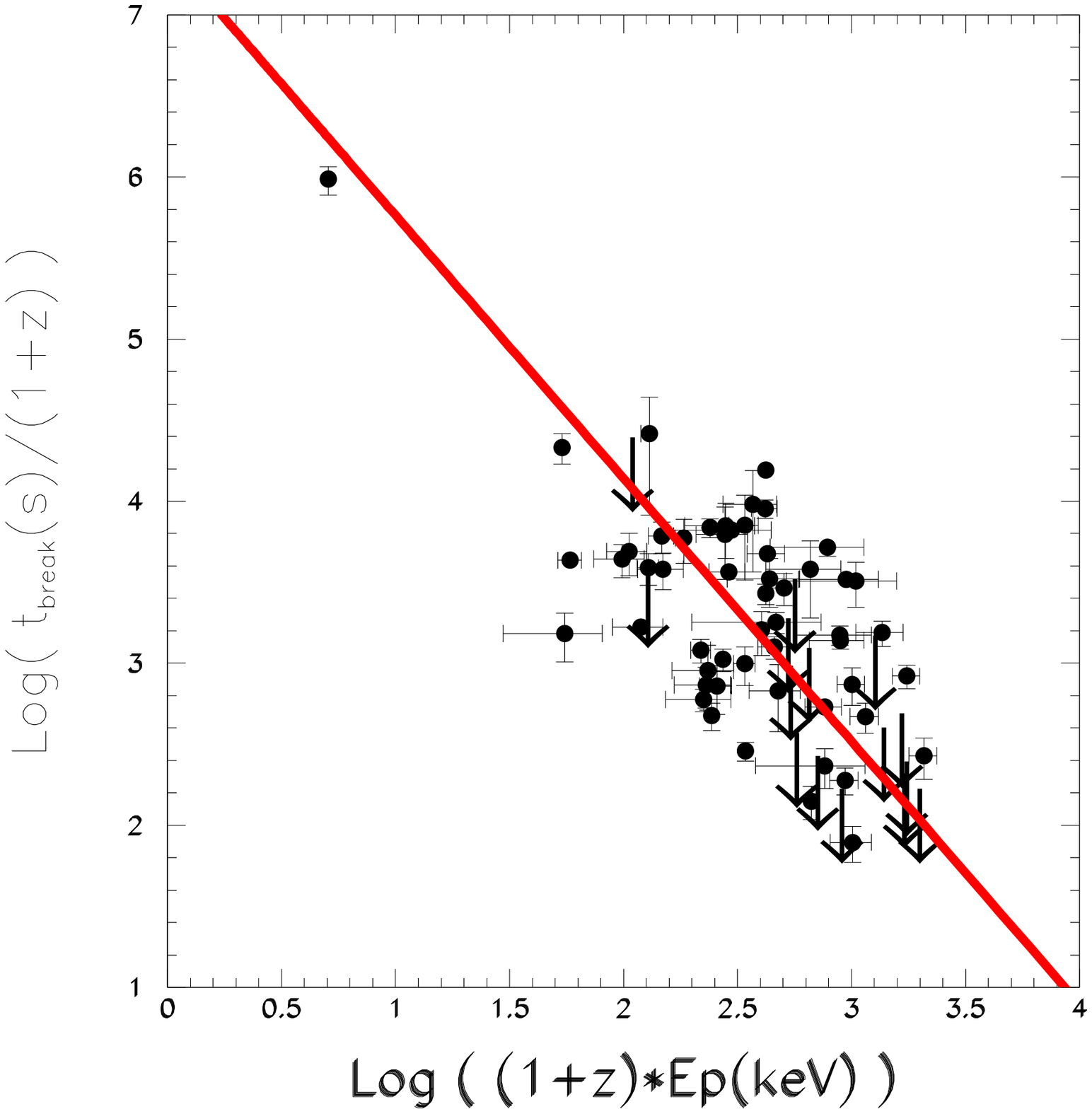,width=16.cm,height=16.cm}
\caption{The binary correlation $t'_b-E'_p$ 
observed in 67 Swift GRBs with measured redshift,
$t_b$, and $E_p$ and its  best fit power-law  
(straight line with a power-law index -1.61). 
Arrows indicate observational upper bounds on early time deceleration 
breaks
before the beginning of the Swift/XRT observations or
hidden under the prompt emission tail.}
\label{FIGC6}
\end{figure} 
\end{document}